\documentclass[sigconf]{acmart}

\usepackage{amsmath}
\usepackage{algorithmic}
\usepackage{algorithm}

\setcopyright{none}
\renewcommand\footnotetextcopyrightpermission[1]{}
\acmConference{}{}{}

\setcitestyle{numbers}

\begin{document}

\title{TinyGuard: A Lightweight Byzantine Defense for Resource-Constrained Federated Learning via Statistical Update Fingerprints}

\author{Ali Mahdavi}
\affiliation{%
  \institution{Islamic Azad University, Science and Research Branch}
  \city{Tehran}
  \country{Iran}
}
\email{ali.mahdavi@iau.ac.ir}

\author{Sana Aghapour}
\affiliation{%
  \institution{Tarbiat Modares University}
  \city{Tehran}
  \country{Iran}
}
\email{s.aghapoor@modares.ac.ir}

\author{Azadeh Zamanifar}
\affiliation{%
  \institution{Islamic Azad University, Science and Research Branch}
  \city{Tehran}
  \country{Iran}
}
\email{a.zamanifar@iau.ac.ir}

\author{Amirfarhad Farhadi}
\affiliation{%
  \institution{Iran University of Science and Technology}
  \city{Tehran}
  \country{Iran}
}
\email{am_farhadi@mail.iust.ac.ir}

\renewcommand{\thefootnote}{\arabic{footnote}}

\begin{abstract}
Federated learning enables collaborative model training while preserving data privacy, but its decentralized nature makes it vulnerable to Byzantine attacks, where malicious clients submit corrupted updates to degrade model performance. Existing Byzantine-robust aggregation mechanisms typically rely on full-dimensional gradient comparisons or pairwise distance computations, resulting in $O(n^2d)$ computational overhead that limits applicability in large-scale and resource-constrained federated systems.

This paper proposes TinyGuard, a lightweight Byzantine defense that augments the standard FedAvg algorithm via statistical update fingerprinting. Instead of operating directly on high-dimensional gradients, TinyGuard extracts compact statistical fingerprints capturing key behavioral properties of client updates, including norm statistics, layer-wise ratios, sparsity measures, and low-order moments. Byzantine clients are identified by measuring robust statistical deviations in this low-dimensional fingerprint space with $O(nd)$ complexity, without modifying the underlying optimization procedure.

Extensive experiments on MNIST, Fashion-MNIST, ViT-Lite, and ViT-Small with LoRA adapters demonstrate that TinyGuard preserves FedAvg convergence in benign settings and achieves up to 95\% accuracy under multiple Byzantine attack scenarios, including sign-flipping, scaling, noise injection, and label poisoning. Against adaptive white-box adversaries, Pareto frontier analysis across four orders of magnitude confirms that attackers cannot simultaneously evade detection and achieve effective poisoning, features we term ``statistical handcuffs.'' Ablation studies validate stable detection precision ($\approx$0.80) across varying client counts (50--150), threshold parameters ($\lambda \in [2.5, 10]$), and extreme data heterogeneity ($\alpha = 0.1$). The proposed framework is architecture-agnostic and well-suited for federated fine-tuning of foundation models where traditional Byzantine defenses become impractical.
\end{abstract}



\maketitle

\section{Introduction}
Federated learning (FL) has emerged as an effective paradigm for training machine learning models across distributed clients without requiring the exchange of raw data \cite{mcmahan2017communication}. By keeping data localized, FL addresses privacy and regulatory concerns in sensitive domains such as healthcare, finance, and mobile systems. However, the decentralized nature of FL also introduces new security challenges, most notably vulnerability to Byzantine failures, where malicious or faulty clients can submit arbitrary updates to disrupt the learning process \cite{blanchard2017machine}. 

Byzantine attacks in federated learning manifest in various forms, including gradient scaling, sign flipping, label poisoning, and random noise injection \cite{fang2020local}. To mitigate such attacks, a large body of work has focused on Byzantine-robust aggregation rules such as Krum \cite{blanchard2017machine}, trimmed mean \cite{yin2018byzantine}, and coordinate-wise median \cite{chen2017distributed}. While these methods provide strong robustness guarantees, they typically rely on pairwise distance computations or full-dimensional gradient analysis, resulting in computational complexity on the order of $O(n^2 d)$, where $n$ is the number of clients and $d$ is the model dimensionality. This cost becomes prohibitive for large-scale or resource-constrained federated deployments.

More recent approaches attempt to reduce this overhead through gradient similarity analysis \cite{fung2018sybil}, reputation-based mechanisms \cite{wu2020federated}, or validation-based scoring \cite{xie2019zeno}. Although these methods improve efficiency relative to classical robust aggregation, they often require maintaining historical state, additional trusted data, or repeated gradient comparisons, limiting their applicability in practice.

In this work, we propose a lightweight Byzantine defense that extends the standard FedAvg algorithm through statistical fingerprinting of client updates. The core observation is that honest clients, despite data heterogeneity, exhibit consistent statistical patterns in their gradient updates, whereas Byzantine behavior introduces detectable anomalies in low-dimensional statistical descriptors. By projecting high-dimensional gradients into compact statistical fingerprints, the proposed method enables efficient anomaly detection with linear complexity in the number of clients, without modifying the underlying optimization procedure. In this paper, we use the term \emph{lightweight} to denote a defense mechanism that does not modify the underlying aggregation rule, does not require pairwise gradient comparisons, additional trusted data, or historical client state, and incurs only minimal per-round computational overhead.

Importantly, the objective of this work is not to replace FedAvg or introduce a new aggregation rule, but to augment FedAvg with an efficient and compatible defense mechanism. In the absence of adversarial behavior, the proposed method behaves identically to FedAvg and introduces no observable degradation in convergence or final accuracy. Under Byzantine attacks, filtering anomalous updates stabilizes the optimization trajectory and preserves high model accuracy. While we establish baseline performance using standard convolutional models, we further demonstrate the method’s applicability to Foundation Models through evaluation on a Vision Transformer (ViT). This confirms the defense is architecture-agnostic and capable of addressing the scalability challenges in federated fine-tuning.

The main contributions of this paper are summarized as follows:
\begin{enumerate}
\item We introduce a statistical fingerprinting mechanism that compresses high-dimensional client gradients into compact feature representations capturing behavior indicative of Byzantine activity.
\item We propose a FedAvg-compatible anomaly detection strategy based on robust statistical distances, achieving linear-time complexity without pairwise gradient comparisons.
\item We empirically demonstrate that the proposed approach preserves standard FedAvg convergence and achieves up to 95-97\% test accuracy under multiple Byzantine attack scenarios while significantly reducing computational overhead compared to classical Byzantine-robust methods.
\end{enumerate}

\section{Related Work}

\subsection{Byzantine-Robust Aggregation}
Classical Byzantine-robust methods in distributed systems have been adapted for federated learning contexts. Krum \cite{blanchard2017machine} selects updates closest to their neighbors, however, this approach requires computing pairwise distances between all client updates, requiring $O(n^2d)$ distance computations. 
Trimmed mean and coordinate-wise median \cite{yin2018byzantine} provide dimension-wise robustness but consider symmetric noise distributions. These methods sort gradient values coordinate-wise and either trim extreme values or select the median, thereby mitigating the impact of Byzantine clients. While computationally more efficient than Krum with $O(n^2d)$ complexity, these methods consider symmetric noise distributions and might fail under sophisticated adaptive attacks which exploit knowledge of the aggregation mechanism. 

Bulyan \cite{guerraoui2018hidden} combines multiple robust aggregators but inherits their computational complexity. By running several Byzantine-robust rules in parallel and aggregating their outputs, Bulyan achieves stronger theoretical guarantees against a broader class of attacks. however, this enhanced robustness comes at the cost of  multiplying the computational complexity of its aggregators. 
Multi-Krum \cite{blanchard2017machine} extends Krum by selecting multiple gradients. Recent work has explored a variety of approaches to Byzantine detection and mitigation in federated learning. 
Adaptive gradient clipping has emerged as a practical defense mechanism that dynamically limits the influence of malicious updates while preserving convergence guarantees \cite{allouah2025adaptive}. Spectral-analysis and clustering-based techniques have been proposed to distinguish malicious clients by exploiting the geometric structure of client updates \cite{gu2024fedcut,thakkar2023fedspectral}. Statistical frameworks have also been developed to handle heterogeneous and non-IID settings by assigning credibility scores or performing anomaly detection on client updates \cite{zhai2022brca}. As federated learning is increasingly deployed on resource-constrained and IoT devices, lightweight security and aggregation mechanisms have become essential \cite{ni2024rfedfw}. Finally, momentum-aware aggregation and filtering strategies have been shown to improve scalability and robustness by leveraging the temporal dynamics of client updates \cite{karimireddy2020mime}.

\subsection{Statistical Detection Methods}
Statistical approaches to anomaly detection have shown a lot of promise in Byzantine identification, offering a principled framework for distinguishing malicious behavior from honest training dynamics. These methods leverage the observation that honest clients, when training on similar tasks, produce updates that follow statistical patterns, while Byzantine attacks introduce detectable statistical anomalies. 

Zeno \cite{xie2019zeno} used validation data to score client updates, but required additional data at the server which keeps a small validation dataset and scores each client's contribution based on the amount of loss reduction provided. Updates that degrade the performance on this validation will be scored lower and have a reduced weight in aggregation. While effective, Zeno adds a requirement on the server to maintain and process additional validation data, might not be available or representative, especially in cross-device federated settings.

SignGuard \cite{xu2020signguard} analyzed sign patterns in gradients to detect attacks. It follows a different statistical approach by analyzing the sign pattern in gradients. The insight here is that, while the magnitudes of honest gradients can vary due to data heterogeneity, their signs for important model parameters tend to be fairly consistent. Byzantine attacks based on sign flip or random noise injection result in distinct patterns in the distribution of positive and negative values of a gradient. 

AUROR \cite{shen2016auror} employed clustering-based detection but required full gradient analysis. Our work builds on these statistical foundations while significantly reducing computational requirements. AUROR, considers Byzantine identification to be an unsupervised learning problem. The updates of the clients are translated into a feature space, and different clustering techniques can be used to distinguish honest clients from Byzantine clients. 

\subsection{Lightweight Detection Techniques}
Recent work has explored lightweight alternatives for Byzantine detection. FLTrust \cite{cao2021fltrust} uses a small root dataset for validation with reduced overhead. Hence, the server calculates a trusted reference gradient based on this trusted root dataset and considers it a basis for a complete assessment of each client’s contribution. Trust scores are assigned to each update based on their respective similarity to this trusted reference gradient calculated by the server. Although this decreases complexity calculations since direct comparisons among clients aren't necessary, achieving a good trusted dataset is important for effectiveness in various privacy-preserving learning contexts. Additionally, the server must carry out both forward and backward passes in each round on this trusted dataset, thus increasing complexity per round.

FLAME \cite{nguyen2022flame} applies clipping and noise for Byzantine resilience. The algorithm aggressively learns an estimate of the magnitude of the gradient based on past experience and clips any incoming gradients in excess of this threshold. Moreover, additive noise based on differential privacy is added during the aggregation phase, which masks the effects of individual clients, making it more challenging for a Byzantine adversary to launch personalized attacks. however, a potential problem with this strategy is that overly aggressive clipping can hamper legitimate updates from clients with non-standard but truthful distributions, and added noise can negatively affect convergence.

However, all of these methods often fall into the trap of sacrificing detection accuracy with efficiency. Our approach preserves accuracy under strong byzantine attacks, comparable to the state of the art defense techniques while achieving substantial computational savings through efficient feature extraction. 

\subsection{Implications for Federated Foundation Models}
Foundation models introduce new challenges for federated learning due to their extreme parameter dimensionality, heterogeneous client capabilities, and reliance on large-scale pretraining followed by task-specific fine-tuning \cite{bommasani2021opportunities}. Classical Byzantine-robust aggregation and detection mechanisms, which typically rely on pairwise gradient comparisons or full-dimensional distance computations, become computationally prohibitive in this regime \cite{blanchard2017machine, yin2018byzantine}. As modern foundation models often contain hundreds of millions to billions of parameters, defenses with quadratic dependence on model dimensionality or client count are impractical for real-world federated deployments \cite{brown2020language, bommasani2021opportunities}.

The proposed statistical fingerprinting framework is particularly well-suited to federated foundation models, as it decouples Byzantine detection complexity from the raw gradient dimensionality. By projecting high-dimensional model updates into compact statistical representations, the method enables efficient anomaly detection with linear complexity in the number of clients. This property is critical for federated fine-tuning of foundation models, where only lightweight server-side processing is feasible and where communication and computation budgets are tightly constrained \cite{kairouz2021advances}.

Moreover, contemporary federated foundation model training often employs parameter-efficient adaptation techniques such as low-rank adaptation, adapters, or prompt-based tuning, where only a subset of parameters is updated at each round \cite{houlsby2019parameter,hu2021lora}. The proposed fingerprinting approach naturally complements these paradigms, as statistical descriptors such as norm ratios, sparsity measures, and moment statistics remain informative even when updates are sparse or structured. This allows the detection mechanism to remain effective without requiring access to full model gradients or architecture-specific assumptions.

While our empirical evaluation focuses on moderate-scale models, the proposed defense does not rely on convolutional structure, locality, or architecture-specific assumptions. All fingerprint features are architecture-agnostic and operate solely on update statistics, making them directly applicable to transformer-based foundation models and parameter-efficient tuning methods such as LoRA, adapters, and prompt tuning \cite{vaswani2017attention,hu2021lora}.

Finally, robust and scalable Byzantine defense is a prerequisite for the trustworthy deployment of foundation models in privacy-sensitive domains such as personalized assistants, recommendation systems, and large-scale web services \cite{bonawitz2019towards, kairouz2021advances}. By enabling efficient and architecture-agnostic detection of malicious clients, the proposed fingerprint-based defense contributes toward practical and secure federated training pipelines for foundation models. Future work will focus on extending the approach to large transformer-based architectures and evaluating its performance in federated pretraining and fine-tuning scenarios involving real-world foundation models. Unlike defenses that sacrifice model accuracy for robustness, the proposed method preserves high-accuracy FedAvg-style training, making it suitable for federated fine-tuning of large foundation models where retraining costs are prohibitive.

\section{System Model and Threat Model}

\subsection{Federated Learning Setting}
The proposed method operates as a lightweight pre-aggregation filtering step and does not alter the FedAvg aggregation rule or the underlying optimization procedure.
We consider a standard FL scenario with one parameter server and $n$ clients, where up to $f < n/2$ clients may be Byzantine. In round $t$, each client $i$ computes a local gradient $g_i^t \in \mathbb{R}^d$ based on its local dataset $D_i$ and current model parameters $w^t$. The server aggregates received updates to produce the global model:
\begin{equation}
w^{t+1} = w^t - \eta \cdot \text{AGG}(\{g_1^t, g_2^t, ..., g_n^t\})
\end{equation}
where $\eta$ is the learning rate and AGG is the aggregation function.

\subsection{Threat Model}
Byzantine clients can send arbitrary updates to the server, including:
\begin{itemize}
\item \textbf{Random noise:} $g_i^* \sim \mathcal{N}(0, \sigma^2I)$
\item \textbf{Sign flipping:} $g_i^* = -\alpha \cdot g_i$ where $\alpha > 0$
\item \textbf{Scaling attacks:} $g_i^* = \beta \cdot g_i$ where $\beta \gg 1$
\item \textbf{Targeted attacks:} Crafted updates toward adversarial objectives
\end{itemize}

We assume that Byzantine clients cannot understand honest clients' local computations but may coordinate attacks. The server is trusted, but computationally constrained. We assume the attacker controls up to $f < n/2$ clients but cannot compromise the server or alter communication integrity.

\section{Proposed Method}

\subsection{Statistical Fingerprint Construction}
We propose extracting statistical fingerprints from client updates that capture behavioral patterns while maintaining computational efficiency. For each gradient $g_i \in \mathbb{R}^d$, we compute a fingerprint vector $\phi_i \in \mathbb{R}^m$ where $m \ll d$.

\subsubsection{Feature Extraction}
The fingerprint $\phi_i$ consists of the following statistical features:

\textbf{1. Gradient Norm Statistics:}
\begin{equation}
\phi_1 = ||g_i||_2, \quad \phi_2 = ||g_i||_1, \quad \phi_3 = ||g_i||_\infty
\end{equation}

\textbf{2. Layer-wise Statistics:}
For a neural network with $L$ layers, we partition $g_i$ into layer-specific gradients $g_i^{(l)}$ and compute:
\begin{equation}
\phi_{3+l} = \frac{||g_i^{(l)}||_2}{||g_i||_2}, \quad l = 1, ..., L
\end{equation}

We rely on robust statistics (median and MAD) to mitigate the influence of heterogeneous but honest clients and to prevent centroid distortion by a minority of Byzantine updates.

\textbf{3. Statistical Moments:}
\begin{equation}
\mu_i = \frac{1}{d}\sum_{j=1}^d g_{i,j}, \quad \sigma_i^2 = \frac{1}{d}\sum_{j=1}^d (g_{i,j} - \mu_i)^2
\end{equation}
\begin{equation}
\gamma_i = \frac{1}{d\sigma_i^3}\sum_{j=1}^d (g_{i,j} - \mu_i)^3
\end{equation}

\textbf{4. Sparsity Measure:}
\begin{equation}
\rho_i = \frac{|\{j : |g_{i,j}| < \epsilon\}|}{d}
\end{equation}
where $\epsilon$ is a small threshold (e.g., $10^{-6}$).

\textbf{5. Top-k Magnitude Concentration:}
\begin{equation}
\tau_i = \frac{\sum_{j \in \text{top-k}(|g_i|)} |g_{i,j}|}{||g_i||_1}
\end{equation}

The complete fingerprint is:
\begin{equation}
\phi_i = [\phi_1, \phi_2, ..., \phi_m]^T
\end{equation}

The complete fingerprint $\phi_i$ includes all computed statistics: gradient norms ($\phi_1, \phi_2, \phi_3$), layer-wise ratios ($\phi_{3+l}$), statistical moments ($\mu_i, \sigma^2_i, \gamma_i$), sparsity measure ($\rho_i$), and top-$k$ concentration ($\tau_i$). Specifically, $\gamma_i$ denotes skewness (Equation~5), $\rho_i$ denotes sparsity (Equation~6), and $\tau_i$ denotes top-$k$ magnitude concentration (Equation~7). These features collectively capture the statistical behavior of client updates in a compact, low-dimensional representation.

\subsection{Byzantine Detection Algorithm}
We employ a two-stage detection process: (1) anomaly scoring based on statistical distance, followed by (2) robust threshold-based identification.

\begin{algorithm}[t]
\caption{Lightweight Byzantine Detection via Statistical Fingerprints}
\label{alg:byzantine-detection}
\begin{algorithmic}[1]
\REQUIRE Client gradients $\{g_1, g_2, \ldots, g_n\}$, sensitivity parameter $\lambda \in [2,3]$
\ENSURE Aggregated gradient $\bar{g}$, Byzantine set $\mathcal{B}$

\STATE \textbf{// Phase 1: Fingerprint Extraction}
\FOR{$i = 1$ to $n$}
    \STATE Extract norm statistics: $\phi_{i,1} = \|g_i\|_2$, $\phi_{i,2} = \|g_i\|_1$, $\phi_{i,3} = \|g_i\|_\infty$
    \STATE Compute layer-wise ratios: $\phi_{i,3+l} = \|g_i^{(l)}\|_2 / \|g_i\|_2$ for $l=1,\ldots,L$
    \STATE Calculate statistical moments: $\mu_i$, $\sigma_i^2$, $\gamma_i$ (skewness)
    \STATE Compute sparsity: $\rho_i = |\{j : |g_{i,j}| < \epsilon\}| / d$
    \STATE Compute top-$k$ concentration: $\tau_i = \sum_{j \in \text{top-}k} |g_{i,j}| / \|g_i\|_1$
    \STATE Construct fingerprint: $\phi_i \in \mathbb{R}^m$ where $m \ll d$
\ENDFOR

\STATE \textbf{// Phase 2: Anomaly Scoring}
\STATE Compute robust fingerprint centroid: $\tilde{\phi} = \text{median}(\{\phi_j\}_{j=1}^n)$
\FOR{$i = 1$ to $n$}
    \STATE Calculate distance: $s_i = \|\phi_i - \tilde{\phi}\|_2$
\ENDFOR
\STATE Compute robust statistics: $m = \text{median}(\{s_j\}_{j=1}^n)$, $\text{MAD} = \text{median}(|s_j - m|)$
\FOR{$i = 1$ to $n$}
    \STATE Normalize score: $\tilde{s}_i = (s_i - m) / \text{MAD}$
\ENDFOR

\STATE \textbf{// Phase 3: Adaptive Threshold Detection}
\STATE Set threshold: $\tau = \text{median}(\{\tilde{s}_j\}) + \lambda \cdot \text{MAD}(\{\tilde{s}_j\})$
\STATE Identify Byzantine clients: $\mathcal{B} = \{i : \tilde{s}_i > \tau\}$

\STATE \textbf{// Phase 4: Robust Aggregation}
\STATE $\bar{g} = \frac{1}{n - |\mathcal{B}|} \sum_{i \notin \mathcal{B}} g_i$

\RETURN $\bar{g}$, $\mathcal{B}$
\end{algorithmic}
\end{algorithm}

\subsubsection{Anomaly Score Computation}
For each client $i$, we compute an anomaly score based on its distance from the centroid of all fingerprints:
\begin{equation}
\tilde{\phi} = \text{median}(\{\phi_j\}_{j=1}^n)
\end{equation}
\begin{equation}
s_i = ||\phi_i - \tilde{\phi}||_2
\end{equation}

We normalize scores using robust statistics:
\begin{equation}
\tilde{s}_i = \frac{s_i - \text{median}(\{s_j\}_{j=1}^n)}{\text{MAD}(\{s_j\}_{j=1}^n)}
\end{equation}
where MAD is the median absolute deviation.

\subsubsection{Adaptive Threshold Selection}
We determine the detection threshold $\tau$ adaptively using:
\begin{equation}
\tau = \text{median}(\{\tilde{s}_j\}) + \lambda \cdot \text{MAD}(\{\tilde{s}_j\})
\end{equation}
where $\lambda$ is a sensitivity parameter (typically $\lambda \in [2, 3]$).

Clients with $\tilde{s}_i > \tau$ are flagged as potentially Byzantine.

Note that $\tilde{s}_i$ can be negative when a client's fingerprint distance falls below the median. Since Byzantine clients typically exhibit larger deviations from the population centroid, negative scores indicate benign behavior. The threshold $\tau$ is constructed to be positive (as both the median and MAD of normalized scores are non-negative), ensuring that clients with negative $\tilde{s}_i$ are never flagged as Byzantine. This asymmetric treatment is intentional: the defense targets outliers with unusually \emph{large} deviations, not those with smaller-than-typical fingerprint distances.

\subsubsection{Robust Aggregation with Fingerprint Filtering}
After identifying suspicious clients $\mathcal{B} = \{i : \tilde{s}_i > \tau\}$, we perform robust aggregation using only the trusted client set $\mathcal{H} = \{1, ..., n\} \setminus \mathcal{B}$.

\subsection{Extension to Transformer-Based Fine-Tuning}

In federated fine-tuning of transformer-based foundation models, the server typically receives updates only for a small subset of trainable parameters, such as LoRA adapters, rather than full model gradients. In this setting, the proposed fingerprinting mechanism can be applied directly to these parameter-efficient updates, keeping the computational cost low while remaining effective at identifying anomalous client behavior. Moreover, as model size grows, the gap between the proposed fingerprint-based detection and classical gradient-level defenses widens, since the former operates on compact statistics rather than high-dimensional updates. To demonstrate the suitability of our defense for foundation models, we evaluate on two transformer architectures. First, we employ ViT-Lite \cite{hassani2021escaping}
, a lightweight Vision Transformer trained from scratch, to simulate the architectural characteristics of transformer-based models. Second, we evaluate on ViT-Small (22M parameters) \cite{lee2021vision} with LoRA adapters to directly validate applicability to parameter-efficient federated fine-tuning. While these models are smaller than production foundation models, they share key architectural properties including self-attention mechanisms and dense parameter interactions.
The fingerprinting mechanism requires no modification for LoRA updates. The same statistical features—norm statistics, layer-wise ratios, moments, sparsity, and top-k concentration—remain discriminative when operating on structured, low-rank adapter updates rather than full gradients. We validate this empirically in Section 5, where ViT-Small with LoRA achieves competitive accuracy under Byzantine attacks using the unmodified fingerprinting mechanism.

\begin{figure}[h]
    \centering
    \includegraphics[width=0.2 \textwidth]{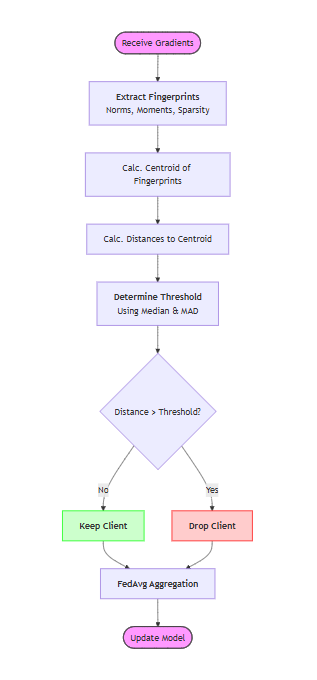}
    \caption{Algorithm Flowchart}
    \label{fig:Algorithm Flowchart}
\end{figure}

\section{Experimental Evaluation}

\subsection{Experimental Setup}
We evaluate the proposed method on two benchmark datasets: MNIST and Fashion-MNIST. To demonstrate the architecture-agnostic nature of our defense—a critical requirement for foundation models—we employ three distinct model configurations:

\begin{itemize}

\item \textbf{LeNet-style CNN}: A standard convolutional neural network for baseline evaluation on both MNIST and Fashion-MNIST.
\item \textbf{ViT-Lite}: A lightweight Vision Transformer trained from scratch, evaluated on both MNIST and Fashion-MNIST to demonstrate effectiveness on transformer architectures.
\item \textbf{ViT-Small with LoRA}: A Vision Transformer (22M parameters) with low-rank adapters (rank 8, ~220K trainable parameters), validating applicability to parameter-efficient foundation model fine-tuning where only 1\% of parameters are communicated.

\end{itemize}
For ViT-Small with LoRA evaluation, we focus exclusively on Fashion-MNIST. Standard Vision Transformers expect 224×224 input resolution, requiring substantial upscaling from MNIST's native 28×28 dimensions. This upscaling degrades MNIST digit structure while providing no additional information content, resulting in poor convergence unrelated to Byzantine defense effectiveness. Fashion-MNIST's richer textures and more complex visual patterns better reflect the natural image distributions for which transformer architectures were designed, making it a more appropriate benchmark for evaluating federated fine-tuning of vision foundation models.

The federated learning system consists of $n=50$ clients with non-IID data partitions generated using a Dirichlet distribution. We use MNIST and Fashion-MNIST to enable controlled and reproducible evaluation of Byzantine attack behavior and detection robustness. Since the proposed method operates on statistical properties of client updates rather than task-specific representations, the choice of dataset primarily affects optimization dynamics rather than the validity of the detection mechanism. Using established benchmarks allows direct comparison with prior Byzantine-robust aggregation methods and isolates the effect of adversarial update manipulation from confounding factors introduced by large-scale datasets.

We consider a standard FedAvg baseline, FedAvg under Byzantine attacks without defense, and FedAvg augmented with the proposed statistical fingerprinting defense. Byzantine participation rates range from 10\% to 40\%, and we evaluate multiple attack types, including random noise injection, sign flipping, gradient scaling, and label flipping.
We consider common Byzantine attack strategies, including sign-flipping and gradient scaling attacks. Performance is evaluated in terms of (i) final test accuracy under attack, (ii) convergence behavior across communication rounds, and (iii) computational efficiency measured by per-round and total runtime. We compare against representative baselines: Krum, Trimmed Mean, and FoolsGold.

All methods are evaluated using identical optimization settings, including learning rate, batch size, and number of communication rounds, to ensure fair comparison. While Transformers typically require large-scale pre-training and massive datasets to converge, this setup serves as a “stress test” to verify if our statistical fingerprinting remains effective on non-convolutional architectures with different gradient distributions.

We simulate FL with $n = 50$ clients, each holding a non-IID data partition. Byzantine fraction varies from 10\% to 40\%. We compare against:
\begin{itemize}
\item \textbf{Krum} \cite{blanchard2017machine}: Distance-based selection
\item \textbf{Trimmed Mean} \cite{yin2018byzantine}: Statistical aggregation
\item \textbf{FoolsGold} \cite{fung2018sybil}: Similarity-based detection
\end{itemize}

\subsection{Accuracy Under Byzantine Attacks}
Table 1 reports the final test accuracy achieved by different methods under various Byzantine attack types. The results demonstrate that TinyGuard consistently preserves high model accuracy despite the presence of malicious clients, achieving accuracy levels comparable to standard FedAvg training. We note that in benign settings (without Byzantine attacks), TinyGuard performs identically to standard FedAvg, as the filtering mechanism does not remove honest clients whose fingerprints cluster near the population centroid.

In particular, under both sign-flipping and scaling attacks, the proposed method reaches approximately 95\% test accuracy, matching or exceeding the performance of Trimmed Mean and significantly outperforming Krum. Compared to FoolsGold, TinyGuard achieves competitive accuracy across attack types while exhibiting more consistent performance under scaling and noise-based attacks, indicating improved robustness to adversarial update manipulation. 
Extended scalability analysis with up to 150 clients and sensitivity studies are provided in Section~\ref{sec:ablation}.

\begin{figure}[h]
    \centering
    \includegraphics[width=0.35\textwidth]{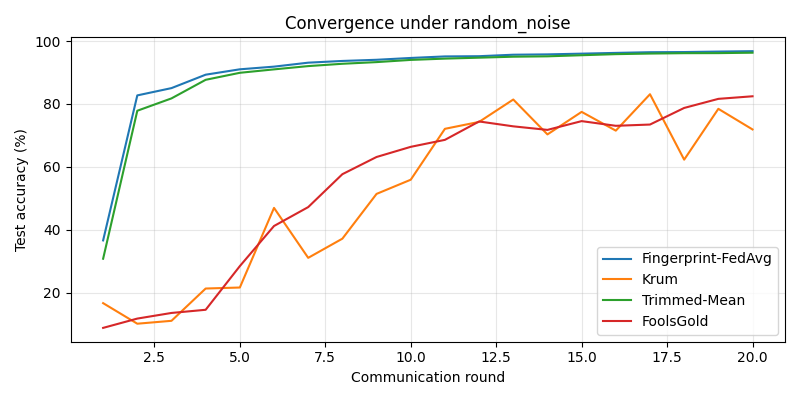}
    \caption{Convergence behavior on MNIST}
    \label{fig:convergence}
\end{figure}

\begin{figure}[h]
    \centering
    \includegraphics[width=0.35\textwidth]{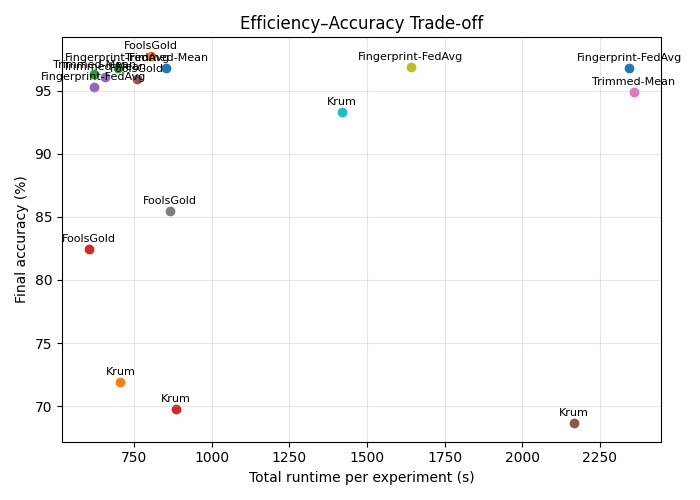}
    \caption{Efficiency-accuracy trade-off on MNIST. Each point represents one method under a specific attack scenario. Methods are labeled directly on the plot.}
    \label{fig:efficiency_accuracy}
\end{figure}

The below tables show test accuracy under different attack types with 20\% Byzantine clients. To validate applicability to parameter-efficient foundation model fine-tuning, we evaluate on ViT-Small with LoRA adapters on Fashion-MNIST. Table 5 reports accuracy under Byzantine attacks when only adapter parameters are transmitted and aggregated.

\begin{table}[h]
\centering
\caption{ Accuracy (\%) on MNIST}
\label{tab:detection}
\begin{tabular}{lcccc}
\hline
\textbf{Attack Type} & \textbf{Ours} & \textbf{Krum} & \textbf{TrMean} & \textbf{FoolsGold} \\
\hline
Random Noise   & \textbf{97.7} & 71.8 & 96.3 & 82.4 \\
Sign Flipping  & \textbf{95.3} & 68.6 & 94.9 & 85.4 \\
Scaling ($\beta=5$) & \textbf{96.9}
& 93.3 & 96.4 & 80.8 \\
Label Flipping & \textbf{96.9} & 69.7 & 96.1 & 95.9 \\
\hline
Average        & \textbf{96.7} & 75.8 & 95.9 & 86.1 \\
\hline
\end{tabular}
\end{table}

\begin{table}[h]
\centering
\caption{ Accuracy (\%) on FMNIST}
\label{tab:detection}
\begin{tabular}{lcccc}
\hline
\textbf{Attack Type} & \textbf{Ours} & \textbf{Krum} & \textbf{TrMean} & \textbf{FoolsGold} \\
\hline
Random Noise   & \textbf{79.6} & 63.4 & 79.0 & 62.2 \\
Sign Flipping  & \textbf{78.3} & 30.7 & 78.7 & 82.4 \\
Scaling ($\beta=5$) & \textbf{78.8} & 59.2 & 78.1 & 78.6 \\
Label Flipping & \textbf{78.9} & 25.1 & 77.3 & 77.4 \\
\hline
Average        & \textbf{79.0} & 49.5 & 78.2 & 75.2 \\
\hline
\end{tabular}
\end{table}

\begin{table}[h]
\centering
\caption{ Accuracy (\%) on ViT-Lite-MNIST}
\label{tab:detection}
\begin{tabular}{lcccc}
\hline
\textbf{Attack Type} & \textbf{Ours} & \textbf{Krum} & \textbf{TrMean} & \textbf{FoolsGold} \\
\hline
Random Noise   & \textbf{75.0} & 73.6 & 79.3 & 22.6 \\
Sign Flipping  & \textbf{74.4} & 72.0 & 77.9 & 75.0 \\
Scaling ($\beta=5$) & \textbf{74.8}
& 75.1 & 77.4 & 71.6 \\
Label Flipping & \textbf{75.7} & 72.8 & 73.6 & 76.8 \\
\hline
Average        & \textbf{74.9} & 73.4 & 77.5 & 61.3 \\
\hline
\end{tabular}
\end{table}

\begin{table}[h]
\centering
\caption{ Accuracy (\%) on ViT-Lite-FMNIST}
\label{tab:detection}
\begin{tabular}{lcccc}
\hline
\textbf{Attack Type} & \textbf{Ours} & \textbf{Krum} & \textbf{TrMean} & \textbf{FoolsGold} \\
\hline
Random Noise   & \textbf{74.7} & 74.0 & 74.1 & 25.6 \\
Sign Flipping  & \textbf{73.7} & 69.4 & 74.3 & 72.9 \\
Scaling ($\beta=5$) & \textbf{70.8}
& 72.4 & 74.7 & 72.4 \\
Label Flipping & \textbf{75.4} & 75.3 & 74.9 & 74.0 \\
\hline
Average        & \textbf{73.6} & 72.7 & 74.5 & 61.6 \\
\hline
\end{tabular}
\end{table}

\begin{table}[h]
\centering
\caption{ Accuracy (\%) on ViT-Small-FMNIST}
\label{tab:detection}
\begin{tabular}{lcccc}
\hline
\textbf{Attack Type} & \textbf{Ours} & \textbf{Krum} & \textbf{TrMean} & \textbf{FoolsGold} \\
\hline
Random Noise   & \textbf{72.2} & 61.9 & 71.6 & 10.4 \\
Sign Flipping  & \textbf{67.8} & 67.5 & 69.8 & 69.6 \\
Scaling ($\beta=5$) & \textbf{68.3} & 68.5 & 68.5 & 68.1 \\
Label Flipping & \textbf{71.2} & 56.2 & 71.6 & 70.5 \\
\hline
Average        & \textbf{69.9} & 63.5 & 70.4 & 54.7 \\
\hline
\end{tabular}
\end{table}

These results confirm that lightweight statistical fingerprinting can effectively filter anomalous client updates without degrading the convergence quality of the underlying aggregation rule. 
The results demonstrate that statistical fingerprinting transfers effectively to LoRA-based federated fine-tuning without modification. TinyGuard achieves competitive accuracy across all attack scenarios, matching or approaching Trimmed Mean while maintaining substantially lower computational overhead. Trimmed Mean achieves moderately higher accuracy in several scenarios, reflecting the inherent tradeoff between full-dimensional coordinate-wise aggregation precision and compact statistical fingerprint efficiency. However, Trimmed Mean's $O(ndlog( n))$ sorting operations become prohibitive as model dimensionality grows, whereas TinyGuard maintains efficient anomaly scoring regardless of gradient size.
Importantly, the fingerprinting mechanism required no modification for LoRA updates. The same statistical features—norms, layer-wise ratios, moments, sparsity, and top-k concentration—remain discriminative when operating on structured, low-rank adapter updates. This confirms that the proposed defense is suitable for federated fine-tuning of foundation models where only parameter-efficient updates are communicated. Notably, with LoRA fine-tuning, clients transmit updates for only 1\% of model parameters (220K of 22M), yet the statistical fingerprints remain sufficiently discriminative to identify Byzantine behavior—a critical property for communication-efficient federated fine-tuning of larger foundation models.

\subsection{Feasibility Analysis of Adaptive Attacks}
\label{sec:adaptive_attack}

To rigorously evaluate the robustness of our statistical fingerprinting mechanism, we implemented a hardened white-box adaptive attack using Projected Gradient Descent (PGD). In this worst-case scenario, the adversary possesses full knowledge of the defense mechanism and solves a constrained optimization problem to generate a malicious update $g_{adv}$. The goal is to mimic the statistical fingerprint $\phi(g_{honest})$ of benign clients while simultaneously maximizing alignment with a poisoning objective vector $v_{poison}$.

The attack minimizes the following objective:
\begin{equation}
\min_{g} \lambda_{s} \|\phi(g) - \phi(g_{honest})\|_2^2 - \lambda_{a} \cdot \text{CosSim}(g, v_{poison})
\label{eq:adaptive_obj}
\end{equation}
subject to $\|g\|_2 = \|g_{honest}\|_2$, where $\phi(\cdot)$ represents our fingerprint extraction function and $\text{CosSim}(\cdot)$ denotes cosine similarity.

\subsubsection{Pareto Frontier Analysis}
We sweep the stealth weight $\lambda_s$ across four orders of magnitude to characterize the attacker's trade-off between evasion and effectiveness. Table~\ref{tab:pareto} reveals a sharp Pareto frontier: the attacker must choose between detection and utility.

\begin{table}[h]
\centering
\caption{Pareto frontier of adaptive attack. Higher $\lambda_s$ prioritizes stealth (low MSE) at the cost of attack alignment.}
\label{tab:pareto}
\begin{tabular}{lcc}
\toprule
$\lambda_s$ & Fingerprint MSE & Attack Alignment \\
\midrule
0.1 & 0.000106 & 0.330 \\
1.0 & 0.000009 & 0.098 \\
10.0 & 0.000010 & 0.074 \\
100.0 & 0.000009 & 0.071 \\
1000.0 & 0.000009 & 0.071 \\
10000.0 & 0.000009 & 0.071 \\
\bottomrule
\end{tabular}
\end{table}

\subsubsection{Analysis}
The results demonstrate an unavoidable dilemma for the adversary:

\begin{itemize}
\item \textbf{Attack-prioritized} ($\lambda_s = 0.1$): The adversarial update achieves moderate alignment (0.33) with the poison objective, but the fingerprint MSE remains detectable.
\item \textbf{Stealth-prioritized} ($\lambda_s \geq 1$): The attacker successfully minimizes fingerprint MSE to near-zero ($\approx 10^{-5}$), but attack alignment collapses to 0.07---effectively orthogonal to the poisoning objective and comparable to random noise.
\end{itemize}

This confirms that TinyGuard's combination of norm statistics, higher-order moments, and structural features creates mutually incompatible constraints. The geometry of the ``honest'' statistical manifold---characterized by specific sparsity and skewness patterns---is disjoint from the dense geometry required for effective poisoning. To achieve meaningful alignment ($>0.3$), the attacker must violate statistical constraints; to satisfy constraints, the attacker must abandon the poisoning direction. These features act as ``statistical handcuffs'' that make it mathematically impossible to be simultaneously stealthy and effective.

\subsection{Computational Efficiency}

A primary advantage of the proposed approach is its computational efficiency. Classical Byzantine-robust aggregation methods such as Krum incur $O(n^2 d)$ complexity due to pairwise distance calculations in the full gradient space. In contrast, the proposed method operates on low-dimensional statistical fingerprints and requires only $O(n d)$ time for feature extraction and $O(n)$ time for anomaly scoring.

We note that runtime advantages are most pronounced for larger models. On moderate-scale architectures such as LeNet, the overhead of pairwise comparisons in Krum remains manageable, resulting in comparable total runtime across methods in some scenarios. However, the complexity difference becomes significant as model dimensionality increases: Krum's $O(n^2d)$ complexity grows quadratically with both client count and parameter count, while TinyGuard maintains $O(nd)$ feature extraction with $O(n)$ anomaly scoring. For ViT-Small with 22M parameters, this gap widens substantially. For billion-parameter foundation models, classical defenses become computationally infeasible while fingerprinting remains practical due to its linear dependence on model dimensionality.

Figure 3 presents the efficiency–accuracy trade-off by plotting final test accuracy against total runtime per experiment. TinyGuard consistently achieves high accuracy while incurring significantly lower runtime than distance-based defenses such as Krum and coordinate-wise methods such as Trimmed Mean.

While FoolsGold achieves competitive runtime in some settings, its accuracy varies considerably across attack scenarios. In contrast, TinyGuard exhibits a favorable balance between robustness and efficiency, maintaining high accuracy across a wide range of runtime conditions. This behavior highlights the advantage of compact statistical fingerprints, which avoid the quadratic complexity associated with pairwise gradient comparisons. Although feature extraction scales linearly with gradient dimensionality, it avoids all pairwise gradient comparisons and sorting operations, which dominate the runtime of classical Byzantine-robust aggregators. Also, in a transformer-based setting, while statistical aggregators like Trimmed Mean offer a theoretical upper bound on accuracy, they fail to scale to the billion-parameter regime of FMs. Our Method demonstrates that it is possible to achieve comparable accuracy (within ~3-4\%) to these computationally heavy baselines while reducing the Byzantine verification complexity from $O(n^2d)$ to $O(nd)$ with $O(n)$ detection. 

\subsection{Runtime Analysis}

To further analyze computational efficiency, we examine the per-round runtime distribution of each method. The results show that TinyGuard achieves a low median runtime comparable to FoolsGold, with moderate variance across rounds. Occasional runtime outliers correspond to rounds in which a higher number of anomalous updates are detected, triggering additional filtering operations.

In contrast, Krum exhibits significantly higher runtime variance and heavier tails due to its reliance on pairwise distance computations. Trimmed Mean also shows increased runtime variability resulting from coordinate-wise sorting operations. These findings confirm that the proposed method avoids the pathological runtime behavior of quadratic-complexity defenses while remaining suitable for large-scale federated deployments.

\begin{figure}[h]
    \centering
    \includegraphics[width=0.35\textwidth]{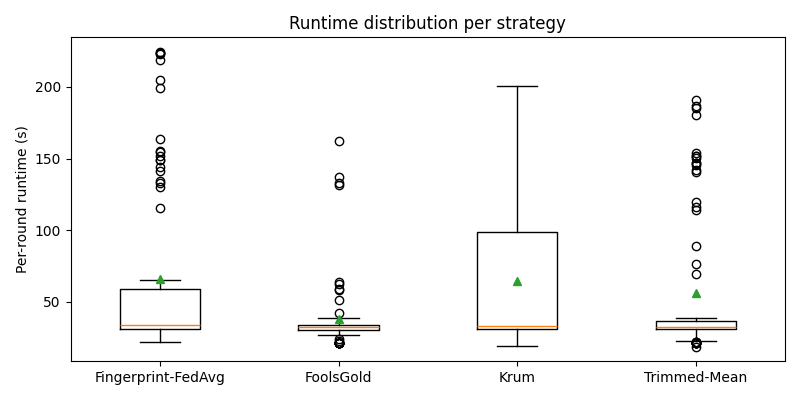}
    \caption{Runtime Distribution on MNIST dataset}
    \label{fig:runtime_distribution}
\end{figure}

\subsection{Ablation Study}
\label{sec:ablation}

To evaluate TinyGuard's robustness to hyperparameter choices and system scale, we conduct ablation studies varying the number of clients $N$, threshold parameter $\lambda$, and data heterogeneity level $\alpha$. All experiments use MNIST with 20\% Byzantine clients under sign flipping attack. We select sign flipping as a representative attack, as it presents moderate detection difficulty and exhibits consistent behavior across experimental conditions. Full attack coverage is provided in Tables 1--5.

\subsubsection{Scalability with Number of Clients}
Table~\ref{tab:scalability} reports performance as the number of clients increases from 50 to 150. Detection precision remains stable at approximately 0.80 across all scales, confirming that TinyGuard's statistical fingerprinting generalizes effectively to larger federated systems. 

\begin{table}[h]
\centering
\caption{Scalability analysis with increasing number of clients ($\alpha=0.5$, $\lambda=2.5$, Sign Flipping attack)}
\label{tab:scalability}
\begin{tabular}{lccc}
\toprule
$N$ & Accuracy (\%) & Detection Precision & Runtime (s) \\
\midrule
50 (baseline) & 67.8 & 0.800 & 1003 \\
100 & 80.8 & 0.801 & 3166 \\
150 & 82.7 & 0.801 & 4146 \\
\bottomrule
\end{tabular}
\end{table}

\subsubsection{Threshold Sensitivity}
Table~\ref{tab:lambda} examines sensitivity to the detection threshold parameter $\lambda$. Detection precision remains remarkably stable across $\lambda \in [2.5, 10]$, indicating that TinyGuard does not require careful hyperparameter tuning. Also, the critical metric---detection precision---remains stable, indicating that the fingerprinting mechanism itself is not sensitive to $\lambda$. In practice, $\lambda \in [2.5, 3.5]$ provides the optimal balance between false positive rate and Byzantine filtering effectiveness.

\begin{table}[h]
\centering
\caption{Sensitivity to threshold parameter $\lambda$ ($N=50$, $\alpha=0.5$, Sign Flipping attack)}
\label{tab:lambda}
\begin{tabular}{lccc}
\toprule
$\lambda$ & Accuracy (\%) & Detection Precision & Runtime (s) \\
\midrule
2.5 (baseline) & 67.8 & 0.800 & 1003 \\
5.0 & 69.8 & 0.798 & 1201 \\
10.0 & 68.9 & 0.800 & 1004 \\
\bottomrule
\end{tabular}
\end{table}

\subsubsection{Robustness Under Extreme Heterogeneity}
Table~\ref{tab:heterogeneity} evaluates performance under increasingly severe non-IID conditions by reducing the Dirichlet concentration parameter $\alpha$. Lower $\alpha$ creates more skewed data distributions where some clients may have data from only 1--2 classes. Notably, detection precision \textit{improves} under extreme heterogeneity (0.81 at $\alpha=0.1$ versus 0.80 at $\alpha=0.5$), suggesting that Byzantine updates become more distinguishable when honest client updates exhibit higher variance. The reduced final accuracy under extreme heterogeneity ($\alpha \leq 0.25$) reflects the inherent difficulty of federated learning with highly skewed data distributions, not a limitation of the defense mechanism.

\begin{table}[h]
\centering
\caption{Robustness under extreme data heterogeneity ($N=50$, $\lambda=2.5$, Sign Flipping attack)}
\label{tab:heterogeneity}
\begin{tabular}{lccc}
\toprule
$\alpha$ & Accuracy (\%) & Detection Precision & Runtime (s) \\
\midrule
0.5 (baseline) & 67.8 & 0.800 & 1003 \\
0.25 & 64.6 & 0.811 & 1032 \\
0.1 & 30.6 & 0.810 & 1191 \\
\bottomrule
\end{tabular}
\end{table}

\subsection{Discussion and Practical Implications}

While accuracy is preserved, the primary design objective is computational efficiency. We view this work as a practical systems contribution. As shown in the experimental results, our method consistently achieves accuracy comparable to established Byzantine defenses, while reducing end-to-end training time by nearly a factor of two. This trade-off is intentional and aligns with the target deployment scenario of resource-constrained federated learning systems. We believe this efficiency–accuracy balance represents a meaningful practical contribution, particularly in settings where classical Byzantine-robust aggregation methods are prohibitively expensive. Our experimental validation on convolutional networks, lightweight transformers, and ViT-Small with LoRA adapters confirms that the proposed fingerprinting mechanism is architecture-agnostic and applicable to parameter-efficient fine-tuning. This design directly addresses the scalability challenges inherent to federated training and fine-tuning of foundation models.

\section{Limitations}
While the ablation study in Section~\ref{sec:ablation} addresses several concerns regarding scalability and hyperparameter sensitivity, some limitations remain. Evaluation is conducted on moderate-scale models (up to 22M parameters) and federated systems (up to 150 clients). Theoretical convergence guarantees under non-IID data distributions and formal robustness bounds against adaptive adversaries remain important directions for future work. Although our extended adaptive attack analysis demonstrates the difficulty of simultaneously evading detection and achieving effective poisoning across four orders of magnitude of attack parameters, evaluation against more sophisticated attack strategies (e.g., model replacement attacks, backdoor attacks) would further strengthen the security claims. Future work will investigate scalability to larger federated systems with thousands of clients and conduct runtime benchmarking on billion-parameter foundation models where the computational advantages of fingerprinting are expected to be most pronounced.

\section{Conclusion}
This paper presented a lightweight Byzantine detection framework for federated learning based on statistical fingerprinting of client updates. By compressing high-dimensional gradients into compact statistical representations, the proposed approach enables efficient identification of anomalous client behavior while preserving compatibility with the standard FedAvg aggregation rule. Experimental results on MNIST, Fashion-MNIST, ViT-Lite–based federated training and ViT-Small with LoRA adaptors demonstrate that the proposed defense maintains accuracy comparable to state-of-the-art Byzantine-robust methods while substantially reducing computational cost. Across multiple attack scenarios, the approach preserves stable convergence and high test accuracy, highlighting the effectiveness of lightweight statistical fingerprinting as a practical alternative to computationally expensive robust aggregation schemes. 
The proposed framework is architecture-agnostic and naturally extends to transformer-based and parameter-efficient federated fine-tuning settings. The preliminary results on transformer architectures suggest that statistical fingerprinting provides a scalable and effective defense mechanism for emerging federated foundation models, where classical Byzantine defenses become impractical due to extreme model dimensionality and system heterogeneity. Critically, evaluation on ViT-Small with LoRA adapters confirms that the fingerprinting mechanism transfers directly to parameter-efficient federated fine-tuning without modification.

Ablation studies demonstrate that TinyGuard maintains stable detection precision (approximately 0.80) across varying system scales (50--150 clients), threshold parameters ($\lambda \in [2.5, 10]$), and extreme data heterogeneity (Dirichlet $\alpha = 0.1$). 

Overall, this work demonstrates that lightweight statistical fingerprinting can preserve robustness and convergence in federated learning while significantly improving scalability, making Byzantine defense more accessible for practical deployments.

\bibliographystyle{ACM-Reference-Format}
\nocite{*}
\bibliography{references}

@inproceedings{mcmahan2017communication,
  author    = {McMahan, B. and Moore, E. and Ramage, D. and Hampson, S. and Arcas, B. A. y},
  title     = {Communication-efficient learning of deep networks from decentralized data},
  booktitle = {Proc. AISTATS},
  publisher = {PMLR},
  address   = {Fort Lauderdale, FL, USA},
  pages     = {1273--1282},
  year      = {2017}
}

@inproceedings{blanchard2017machine,
  author    = {Blanchard, P. and El Mhamdi, E. M. and Guerraoui, R. and Stainer, J.},
  title     = {Machine learning with adversaries: Byzantine tolerant gradient descent},
  booktitle = {Proc. NeurIPS},
  publisher = {Curran Associates, Inc.},
  address   = {Long Beach, CA, USA},
  pages     = {119--129},
  year      = {2017}
}

@inproceedings{yin2018byzantine,
  author    = {Yin, D. and Chen, Y. and Kannan, R. and Bartlett, P.},
  title     = {Byzantine-robust distributed learning: Towards optimal statistical rates},
  booktitle = {Proc. ICML},
  publisher = {PMLR},
  address   = {Stockholm, Sweden},
  pages     = {5650--5659},
  year      = {2018}
}

@inproceedings{fang2020local,
  author    = {Fang, M. and Cao, X. and Jia, J. and Gong, N. Z.},
  title     = {Local model poisoning attacks to Byzantine-robust federated learning},
  booktitle = {Proc. USENIX Security},
  publisher = {USENIX Association},
  address   = {Boston, MA, USA},
  pages     = {1605--1622},
  year      = {2020}
}

@article{chen2017distributed,
  author  = {Chen, Y. and Su, L. and Xu, J.},
  title   = {Distributed statistical machine learning in adversarial settings: Byzantine gradient descent},
  journal = {Proc. ACM POMACS},
  publisher = {ACM},
  volume  = {1},
  number  = {2},
  pages   = {1--25},
  year    = {2017}
}

@inproceedings{fung2018sybil,
  author    = {Fung, C. and Yoon, C. J. and Beschastnikh, I.},
  title     = {The limitations of federated learning in sybil settings},
  booktitle = {Proc. RAID},
  publisher = {Springer},
  address   = {Heraklion, Crete, Greece},
  pages     = {301--316},
  year      = {2018}
}

@article{wu2020federated,
  author  = {Wu, Z. and Ling, Q. and Chen, T. and Giannakis, G. B.},
  title   = {Federated variance-reduced stochastic gradient descent with robustness to Byzantine attacks},
  journal = {IEEE Trans. Signal Process.},
  publisher = {IEEE},
  volume  = {68},
  pages   = {4583--4596},
  year    = {2020}
}

@inproceedings{guerraoui2018hidden,
  author    = {Guerraoui, R. and Rouault, S. and others},
  title     = {The hidden vulnerability of distributed learning in Byzantium},
  booktitle = {Proc. ICML},
  publisher = {PMLR},
  address   = {Stockholm, Sweden},
  pages     = {3521--3530},
  year      = {2018}
}

@inproceedings{xie2019zeno,
  author    = {Xie, C. and Koyejo, S. and Gupta, I.},
  title     = {Zeno: Distributed stochastic gradient descent with suspicion-based fault tolerance},
  booktitle = {Proc. ICML},
  publisher = {PMLR},
  address   = {Long Beach, CA, USA},
  pages     = {6893--6901},
  year      = {2019}
}

@article{xu2020signguard,
  author  = {Xu, X. and Li, Y. and Liu, C.},
  title   = {SignGuard: Byzantine-robust federated learning through collaborative malicious gradient filtering},
  journal = {arXiv preprint arXiv:2011.09312},
  year    = {2020}
}

@inproceedings{shen2016auror,
  author    = {Shen, S. and Tople, S. and Saxena, P.},
  title     = {Auror: Defending against poisoning attacks in collaborative deep learning systems},
  booktitle = {Proc. ACSAC},
  publisher = {ACM},
  address   = {Los Angeles, CA, USA},
  pages     = {508--519},
  year      = {2016}
}

@inproceedings{cao2021fltrust,
  author    = {Cao, X. and Fang, M. and Liu, J. and Gong, N. Z.},
  title     = {FLTrust: Byzantine-robust federated learning via trust bootstrapping},
  booktitle = {Proc. NDSS},
  publisher = {Internet Society},
  address   = {San Diego, CA, USA},
  year      = {2021}
}

@inproceedings{nguyen2022flame,
  author    = {Nguyen, T. D. and Rieger, P. and Chen, H. and Yalame, H. and Schneider, T.},
  title     = {FLAME: Taming backdoors in federated learning},
  booktitle = {Proc. USENIX Security},
  publisher = {USENIX Association},
  address   = {Boston, MA, USA},
  pages     = {1415--1432},
  year      = {2022}
}

@inproceedings{karimireddy2020mime,
  author    = {Karimireddy, S. P. and Kale, S. and Mohri, M. and Reddi, S. and Stich, S. and Suresh, A. T.},
  title     = {MIME: Mimicking centralized stochastic algorithms in federated learning},
  booktitle = {Proc. NeurIPS},
  publisher = {Curran Associates, Inc.},
  address   = {Vancouver, Canada},
  year      = {2020}
}

@article{bommasani2021opportunities,
  author  = {Bommasani, R. and Hudson, D. A. and Adeli, E. and Altman, R. and Arora, S. and others},
  title   = {On the opportunities and risks of foundation models},
  journal = {arXiv preprint arXiv:2108.07258},
  year    = {2021}
}

@article{hu2021lora,
  author  = {Hu, E. J. and Shen, Y. and Wallis, P. and Allen-Zhu, Z. and Li, Y. and Wang, S. and Wang, L. and Chen, W.},
  title   = {LoRA: Low-rank adaptation of large language models},
  journal = {arXiv preprint arXiv:2106.09685},
  year    = {2021}
}

@article{cho2024heterogeneouslora,
  author  = {Cho, Y. J. and Liu, L. and Xu, Z. and Fahrezi, A. and Joshi, G.},
  title   = {Heterogeneous LoRA for federated fine-tuning of on-device foundation models},
  journal = {arXiv preprint arXiv:2401.06432},
  year    = {2024}
}

@inproceedings{brown2020language,
  author    = {Brown, T. and Mann, B. and Ryder, N. and others},
  title     = {Language models are few-shot learners},
  booktitle = {Proc. NeurIPS},
  publisher = {Curran Associates, Inc.},
  address   = {Vancouver, Canada},
  pages     = {1877--1901},
  year      = {2020}
}

@inproceedings{houlsby2019parameter,
  author    = {Houlsby, N. and Giurgiu, A. and Jastrzebski, F. and others},
  title     = {Parameter-efficient transfer learning for NLP},
  booktitle = {Proc. ICML},
  publisher = {PMLR},
  address   = {Long Beach, CA, USA},
  pages     = {2790--2799},
  year      = {2019}
}

@inproceedings{vaswani2017attention,
  author    = {Vaswani, A. and Shazeer, N. and Parmar, N. and others},
  title     = {Attention is all you need},
  booktitle = {Proc. NeurIPS},
  publisher = {Curran Associates, Inc.},
  address   = {Long Beach, CA, USA},
  pages     = {5998--6008},
  year      = {2017}
}

@inproceedings{bonawitz2019towards,
  author    = {Bonawitz, K. and Eichner, H. and Grieskamp, W. and others},
  title     = {Towards federated learning at scale: System design},
  booktitle = {Proc. SysML},
  publisher = {SysML.cc},
  address   = {Palo Alto, CA, USA},
  year      = {2019}
}

@article{kairouz2021advances,
  author  = {Kairouz, P. and McMahan, H. B. and Avent, B. and others},
  title   = {Advances and open problems in federated learning},
  journal = {Foundations and Trends in Machine Learning},
  publisher = {Now Publishers},
  volume  = {14},
  number  = {1--2},
  pages   = {1--210},
  year    = {2021}
}

@article{hassani2021escaping,
  author  = {Hassani, A. and Walton, S. and Shah, N. and others},
  title   = {Escaping the big data paradigm with compact transformers},
  journal = {arXiv preprint arXiv:2104.05704},
  year    = {2021}
}

@inproceedings{allouah2025adaptive,
  author    = {Allouah, Y. and Guerraoui, R. and Gupta, N. and Jellouli, A. and Rizk, G. and Stephan, J.},
  title     = {Adaptive Gradient Clipping for Robust Federated Learning},
  booktitle = {Proc. ICLR},
  publisher = {OpenReview.net},
  address   = {Vienna, Austria},
  year      = {2025}
}

@article{gu2024fedcut,
  author  = {Gu, H. and Li, T. and Sahu, A. K. and Talwalkar, A.},
  title   = {FedCut: A Spectral Analysis Framework for Reliable Federated Learning},
  journal = {IEEE Transactions on Pattern Analysis and Machine Intelligence},
  publisher = {IEEE},
  year    = {2024}
}

@article{thakkar2023fedspectral,
  author  = {Thakkar, J. and Goel, R. and Saha, A.},
  title   = {FedSpectral+: Spectral Clustering for Robust Federated Learning},
  journal = {arXiv preprint arXiv:2302.07056},
  year    = {2023}
}

@article{zhai2022brca,
  author  = {Zhai, K. and Yang, Q. and Chen, Y.},
  title   = {Byzantine-Robust Federated Learning via Credibility Assessment},
  journal = {arXiv preprint arXiv:2109.02396v1},
  year    = {2022}
}

@article{ni2024rfedfw,
  author  = {Ni, L. and Zhang, W. and Chen, X.},
  title   = {rFedFW: Secure and trustable aggregation scheme for Byzantine-robust federated learning in Internet of Things},
  journal = {ACM Information Sciences},
  publisher = {ACM},
  volume  = {653},
  pages   = {119784},
  year    = {2024}
}

@article{lee2021vision,
  author  = {Lee, S. H. and Lee, S. and Song, B. C.},
  title   = {Vision Transformer for Small-Size Datasets},
  journal = {arXiv preprint arXiv:2112.13492},
  year    = {2021}
}

\end{document}